\documentclass[]{interact}

\usepackage{epstopdf}
\usepackage[caption=false]{subfig}
\usepackage{graphicx}
\usepackage{amssymb}
\usepackage{amsmath}
\usepackage{tabularx}
\usepackage{enumitem}
\usepackage{booktabs}
\usepackage{xcolor}
\usepackage{array}
\newcolumntype{?}{!{\vrule width 1pt}}
\DeclareMathOperator*{\argmax}{argmax}
\usepackage{natbib}
\bibpunct[, ]{(}{)}{;}{a}{}{,}

\theoremstyle{plain}

\theoremstyle{definition}

\theoremstyle{remark}

\begin{document}

\articletype{JOURNAL ARTICLE PREPRINT}

\title{Automatic Estimation of Ulcerative Colitis Severity from Endoscopy Videos using Ordinal Multi-Instance Learning\textsuperscript{$\dagger$}\thanks{\textsuperscript{$\dagger$}Accepted for publication in the Special Issue Journal of Computer Methods in Biomechanics and Biomedical Engineering: Imaging \& Visualization and presented at the MICCAI 2021 AE-CAI $|$ CARE $|$ OR 2.0 Workshop.}}

\author{
\name{Evan Schwab and Gabriela Oana Cula and Kristopher Standish and Stephen S. F. Yip and Aleksandar Stojmirovic and Louis Ghanem\textsuperscript{*} and Christel Chehoud\textsuperscript{*}\thanks{\textsuperscript{*}Co-Senior Authors}}
\affil{Janssen R\&D Data Science Analytics and Insights}\thanks{CONTACT Evan Schwab. Email: eschwab1@its.jnj.com} 
}

\maketitle

\begin{abstract}
Ulcerative colitis (UC) is a chronic inflammatory bowel disease characterized by relapsing inflammation of the large intestine. The severity of UC is often represented by the Mayo Endoscopic Subscore (MES) which quantifies mucosal disease activity from endoscopy videos. In clinical trials, an endoscopy video is assigned an MES based upon the most severe disease activity observed in the video. For this reason, severe inflammation spread throughout the colon will receive the same MES as an otherwise healthy colon with severe inflammation restricted to a small, localized segment. Therefore, the extent of disease activity throughout the large intestine, and overall response to treatment, may not be completely captured by the MES. In this work, we aim to automatically estimate UC severity for each frame in an endoscopy video to provide a higher resolution assessment of disease activity throughout the colon. Because annotating severity at the frame-level is expensive, labor-intensive, and highly subjective, we propose a novel weakly supervised, ordinal classification method to estimate frame severity from video MES labels alone. Using clinical trial data, we first achieved $0.92$ and $0.90$ AUC for predicting mucosal healing and remission of UC, respectively. Then, for severity estimation, we demonstrate that our models achieve substantial Cohen's Kappa agreement with ground truth MES labels, comparable to the inter-rater agreement of expert clinicians. These findings indicate that our framework could serve as a foundation for novel clinical endpoints, based on a more localized scoring system, to better evaluate UC drug efficacy in clinical trials.

\end{abstract}
\section{Introduction}
\label{sec:intro}
\begin{figure}[t]
    \centering
       \subfloat[Normal (MES 0)]{\includegraphics[trim={0 12 0 0}, clip, width=0.25\textwidth]{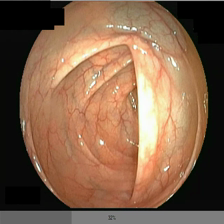}}
   \subfloat[Mild (MES 1)]{\includegraphics[trim={0 12 0 0}, clip, width=0.25\textwidth]{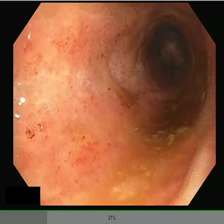}}
    \subfloat[Moderate (MES 2)]{\includegraphics[trim={0 12 0 0}, clip, width=0.25\textwidth]{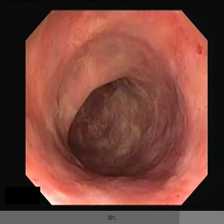}}
         \subfloat[Severe (MES 3)]{\includegraphics[trim={0 12 0 0}, clip, width=0.25\textwidth]{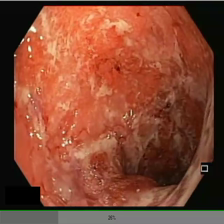}}
    \caption{Example endoscopy video frames and corresponding Mayo Endoscopic Score (MES) of UC severity.}\label{fig:ucseverity}
\end{figure}

Ulcerative colitis (UC) is a disabling and chronic inflammatory bowel disease (IBD) characterized by relapsing inflammation and ulceration of the large intestinal mucosa. Clinical trials in IBD use standardized scoring systems to assess both clinical outcomes and changes in disease activity. One commonly used disease severity score for UC is the total Mayo Score \citep{schroeder1987coated}, which combines clinical disease features, physician global assessment, and mucosal disease burden as determined by video endoscopy. An important component of the total Mayo Score is the Mayo Endoscopic Subscore (MES) which is used to define patient-level UC severity from endoscopic videos and is graded on a discrete 0-3 scale: Normal (MES 0), Mild (MES 1), Moderate (MES 2), Severe (MES 3). See Fig.~\ref{fig:ucseverity} for representative video frames.

\begin{figure}[h]
\centering
\includegraphics[trim={0 0 0 0}, clip, width=.9\textwidth]{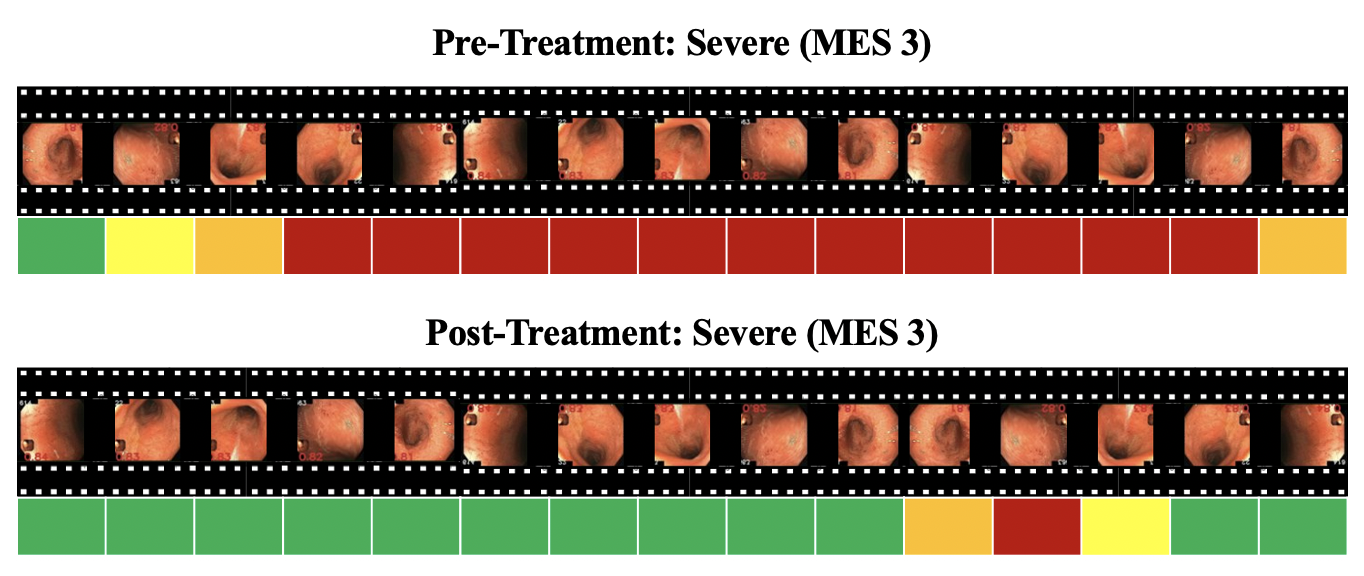}
    \caption{
Human central readers in UC clinical trials assign a video-level MES based on the most severe inflammation observed in an entire endoscopy video.  Therefore, as shown in this illustration, a patient will be scored an MES 3 before and after treatment, without sensitivity to wide-spread improvements (i.e., red (MES 3) reduces to green (MES 0), yellow (MES 1) or orange (MES 2)). This example demonstrates how the current clinical scoring system fails to capture important treatment effects from endoscopy videos and the need to measure severity at a higher resolution.}
    \label{fig:preposttreatment}
\end{figure}

The MES scoring system, as deployed in clinical trials and used in clinical practice, has two notable limitations. First, gastroenterologists (GIs), often serving as central readers in a clinical trial, are required to assign a single video-level MES based upon the most severe disease activity observed in the video. 
Therefore, as illustrated in Fig.~\ref{fig:preposttreatment}, if a patient with severe and widespread UC (MES 3) has marked improvement after treatment, except for a highly localized severe disease activity, that patient will still receive an MES 3. A consequence of this ambiguity is the relative insensitivity of the MES to quantify meaningful, local therapeutic effects of drug treatments which could negatively impact drug development and medical decision making in the clinic.



The second notable limitation is the relatively high subjectivity of scoring UC severity by expert GIs. A Fleiss Kappa ($\kappa$) metric is commonly utilized to assesses inter-rater agreement between multiple raters with the following scale: $0.01–0.20$ (slight), $0.21–0.40$ (fair), $0.41–0.60$ (moderate), $0.61–0.80$ (substantial), and $0.81–1.00$ (nearly perfect). 
As reported in \cite{daperno2014inter}, the inter-rater agreement of video-level MES between 14 expert raters was moderate with a Fleiss $\kappa\!=\!0.53$ ($95\%$ confidence interval (CI) 0.47–0.56). Likewise, \cite{principi2020inter} report moderate Fleiss $\kappa\!=\!0.53$ ($95\%$ CI 0.39–0.66) with 13 expert raters. A Fleiss $\kappa\!=\!0.60$ ($95\%$ CI 0.51-0.69) was also reported between 4 expert raters for the clinical trial data \citep{sands2019ustekinumab} that served as our ground truth labels for training and validation. 

These two limitations, a lack of granularity and moderate inter-rater agreement, motivate the development of an automatic disease activity scoring system for UC which uses frame-level severity scores to provide a more precise measurement of disease distribution and severity throughout the bowel. Such an approach would allow for finer assessments of meaningful therapeutic effects in IBD clinical trials.

Prior work have examined the automatic estimation of frame-level IBD severity scores from endoscopy videos. In particular, \cite{StidhamJAMA19endo,ozawa2019novel,YaoGE20endo} solve supervised learning problems that require frame-level labels from trained experts.  
Employing a set of trained experts to perform frame-level severity labeling, however, is a very time-consuming and expensive task, requiring large numbers of labeled frames suitable for training deep neural networks. In Sec.~\ref{sec:results}, we report that four highly experienced GI clinicians required 3-4 hours per person to label only 200 frames of our dataset. 

Therefore, to overcome the difficulties of frame-level labeling, we utilize weakly supervised deep learning to learn frame-level estimations trained only with video-level labels. In particular, we combine the weakly supervised algorithm multi-instance learning (MIL), with the task of ordinal classification to arrive at a novel ordinal MIL framework for frame-level severity estimation. To the best of our knowledge, this approach has not been previously developed. Most closely related to our work is \cite{gutierrez2021training}, which utilizes a similar form of weak supervision but does not attempt ordinal classification. Additionally, \cite{MohammedCVIU20endo} solves an alternative multi-class disease detection problem on annotated video segments.

This paper is organized as follows: In Sec.~\ref{sec:mil} we introduce the standard MIL framework and build upon this to derive our novel ordinal MIL in Sec~\ref{sec:ordinalclassification}. Then, in Sec.~\ref{sec:experiments} we describe our dataset, pre-processing, training, and experimental setup. We present our results on clinical trial data in Sec.~\ref{sec:results}.


\section{Methods}
\label{sec:methods}
In this section, we combine MIL with ordinal classification to estimate ordinal categories of UC severity at the frame-level using only video-level labels. In Sec.~\ref{sec:mil} we introduce the standard MIL algorithm, traditionally developed for binary classification, which we apply to solve two clinically relevant binary classification tasks. We then extend MIL to the case of ordinal classification for UC severity estimation in Sec.~\ref{sec:ordinalclassification} and compare two established ordinal classification frameworks: 1) an aggregation of an ensemble of ranked binary classifiers and 2) regression with discrete ordinal labels. 

\subsection{Multi-Instance Learning Applied to Endoscopy Videos}
\label{sec:mil}
Multi-instance learning (MIL) is a framework for solving weakly labeled problems in which labeled data can be viewed as a collection of smaller unlabeled parts (instances) that we are interested in estimating. An example medical application is the localization of disease pathology in medical images. While ground truth disease labels (eg. disease vs. no disease) may be relatively easy to provide at the image-level, annotating the specific location of disease pathology in an image has proven to be a more time consuming and difficult undertaking. MIL has been applied by dividing a labeled image into smaller unlabeled image patches and estimating the presence or absence of disease in those patches using weak supervision. 

MIL has been traditionally formulated for binary classification, exploiting the assumption that for images with positive disease labels, \textit{at least one} image patch will be positive, which explains the positive label of the image. Conversely, \textit{all} image patches from a negative image must be negative. This framework has shown effective performance across image modalities like digital histopathology \citep{CampanellaNature19path} and chest X-ray \citep{SchwabISBI20cxr}. Multi-class formulations have also been explored \citep{RemediosMI20ct, mahmood2021histopath}.

In our setting, we apply the same weakly supervised paradigm by substituting videos for images and video frames for image patches. Our first application can be posed as two clinically relevant binary classification tasks: 1) MES 0 vs. (1,2,3) to detect the presence of UC, and 2) MES (0,1) vs. (2,3) to detect mucosal healing which is associated with improved long-term patient outcomes \citep{TurnerGastro20endo}.

Using a neural network, MIL first estimates a probabilistic score for each frame, and then compares an aggregate of those scores with the video label in the loss function. The most typical aggregation is the maximum operator. It effectively selects the frame with the maximum score as representative of the entire video. The max aggregation is well suited for our problem because it mimics the established clinical framework of reporting the maximum MES observed in the video. This framework also allows us to model frames as independent samples, utilizing a classical MIL setting without more advanced temporal modeling.

\begin{figure}[t]
\centering
\includegraphics[trim={0 0 5 5}, clip, width=\textwidth]{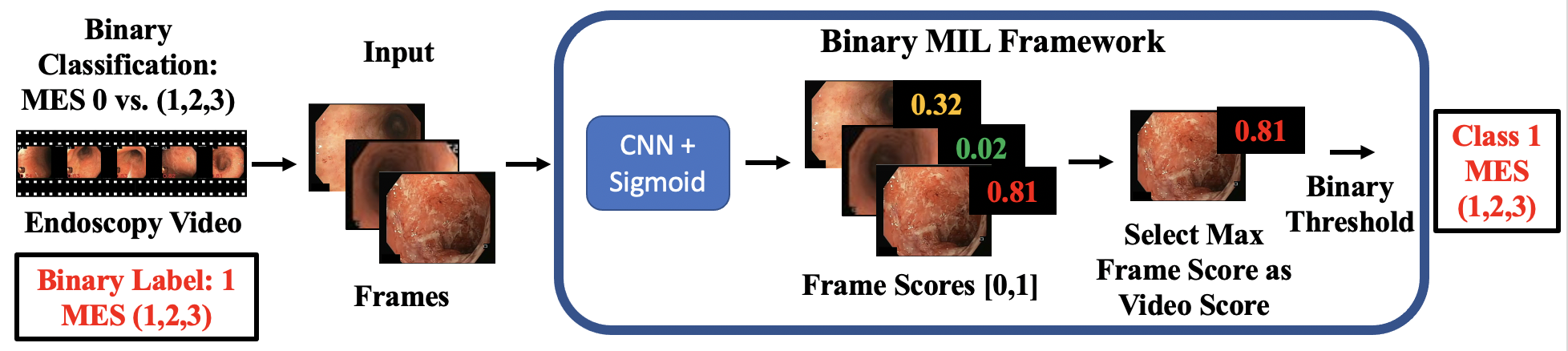}
    \caption{Binary MIL Framework Overview. Endoscopy videos are split into frames and fed into a CNN to generate frame-level probability scores. The frame with the maximum score is selected as the representative for the video and a binary threshold is used to produce a class for the video and each frame.}
    \label{fig:binarymil}
\end{figure}

More specifically, let $V$ be an endoscopic video with $F$ image frames $\{f_i\}_{i=1}^{F}$ and binary label $y = \{0,1\}$. Because we have image data, we use a convolutional neural network (CNN), $\phi$, to estimate continuous disease probability scores for each frame, $p_i = \phi (f_i) \in [0,1]$. Then, to obtain the video-level probability, $p_v$, we select the frame with the maximum score as the representative for the video, i.e. $f_v$ with $v = \argmax_i p_i$, and present this to the binary cross-entropy loss function during training. Applying a threshold $t\in [0,1]$ to $p_v$ produces the video class prediction $\hat{y} \in \{0,1\}$ and at the same time, class predictions $\hat{y}_i\in \{0,1\}$ for each frame in the video. Fig.~\ref{fig:binarymil} provides an overview of MIL for binary classification.  In the next section we extend the standard binary MIL formulation to the case of ordinal classification.  

\subsection{Ordinal MIL for Frame-Level Severity Estimation}
\label{sec:ordinalclassification}
Ordinal classification is appropriate for ordered data labels without an apparent numerical distance between them (i.e., Normal $<$ Mild $<$ Moderate $<$ Severe)\footnote{For simplicity, and by clinical convention, we choose to represent these labels with numerical values (i.e., $0\!<\!1\!<\!2\!<\!3$), but the numerical distance between labels need not be equal.}. Unlike multi-class problems, where misclassifications are weighted equally, misclassifying Severe as Normal should be penalized more than misclassifying Severe as Moderate. Next, we derive the novel combination of MIL with two established ordinal classification frameworks.

\begin{figure}
\centering
\includegraphics[trim={0 0 0 0}, clip, width=\textwidth]{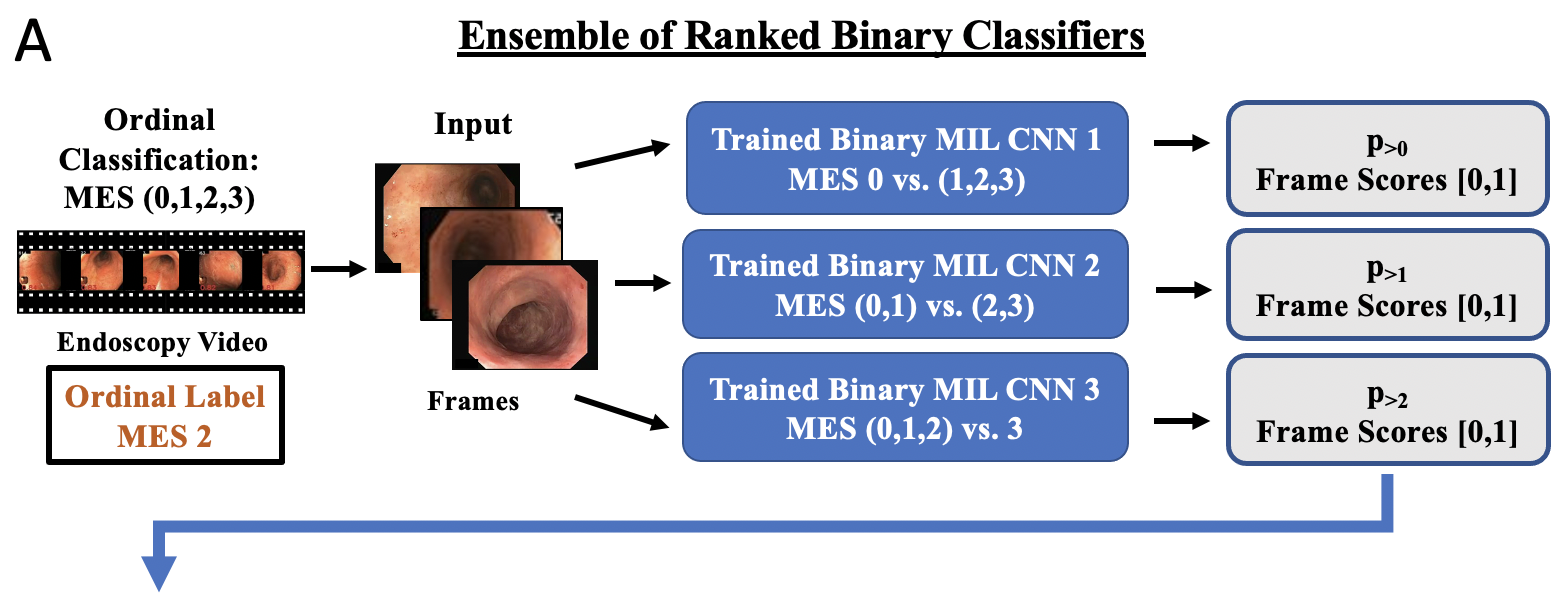}
\includegraphics[trim={0 0 0 0}, clip, width=\textwidth]{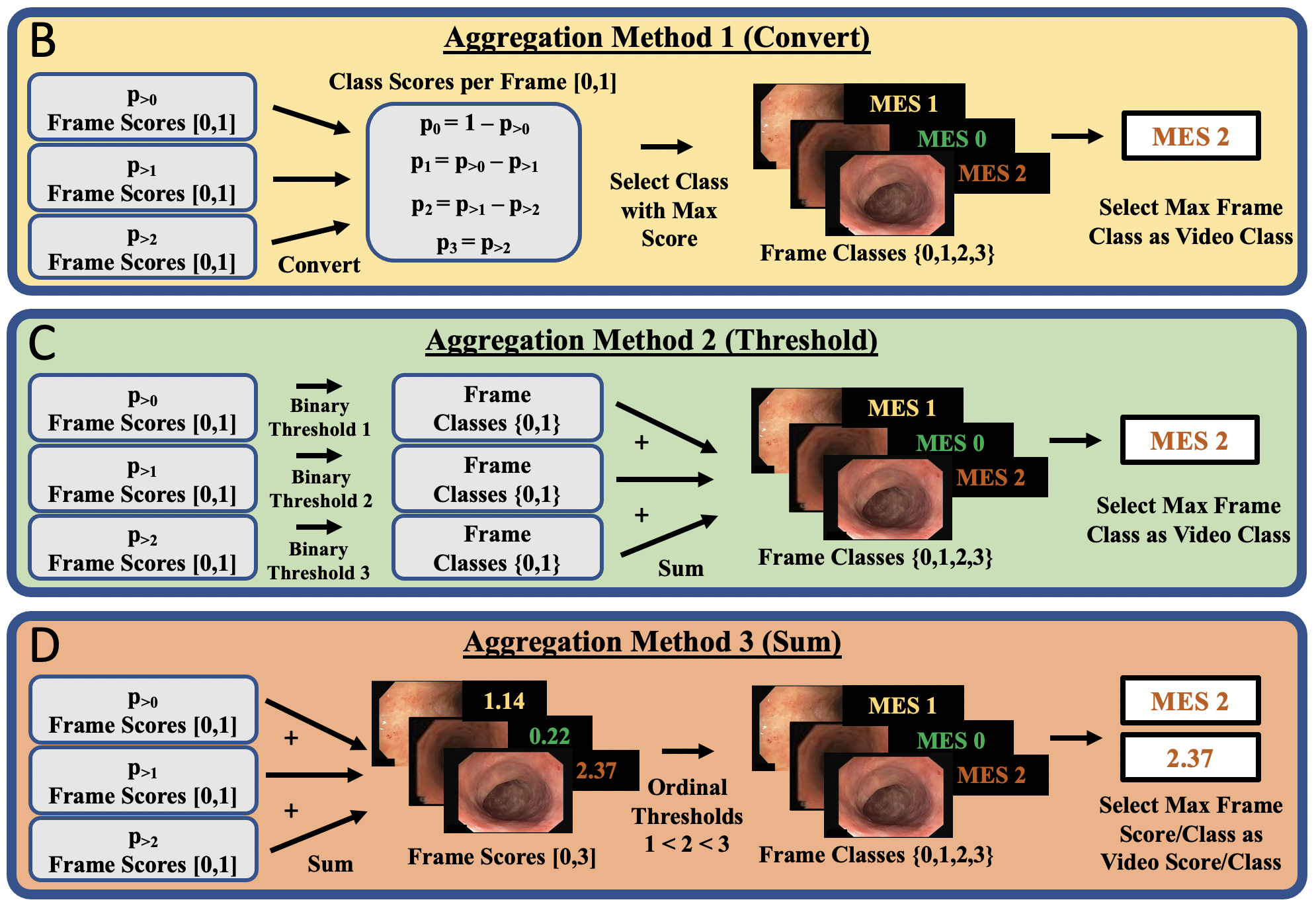}

     \caption{Overview of proposed ordinal MIL framework using an ensemble of ranked binary classifiers (A): $\phi_{>0}$: 0 vs $(1,2,3)$, $\phi_{>1}$: $(0,1)$ vs $(2,3)$, $\phi_{>2}$: $(0,1,2)$ vs $3$. The output of (A) is three probability scores per frame $p_{>0}$, $p_{>1}$, $p_{>2} \in [0,1]$.  We experiment with three different aggregation methods, (B), (C), (D), of the frame probability scores. In (B), $p_{>0}, p_{>1}$ and $p_{>2}$ are converted into multi-class probabilities $p_0$, $p_1$, $p_2$, $p_3 \in [0,1]$, each corresponding to the respective MES class. The maximum probability is selected as the ordinal class for each frame. In (C), three distinct binary thresholds are applied to $p_{>0}$, $p_{>1}$, and $p_{>3}$ resulting in three binary classes per frame. These are summed to generate ordinal classes per frame. In (D), $p_{>0}$, $p_{>1}$, and $p_{>2}$ are first summed to create a continuous score in $[0,3]$ per frame. A set of ordinal thresholds are then applied to generate ordinal classes per frame. In each method, the frame with the maximum class is selected as the video class.}\label{fig:ordinalmil}
\end{figure}

\subsubsection{Ensemble of Ranked Binary Classifiers}
\label{sec:ensemble}
One approach to ordinal classification is to restructure a problem with $M$ ordinal labels as an ensemble of $M\!-\!1$ ranked binary classification tasks and aggregate each output to estimate ordinal classes. Within a traditional fully supervised framework, \cite{Frank01ordinal} convert the $M\!-\!1$ binary probabilities to $M$ multi-class probabilities and take the maximum as the ordinal class. Alternatively, \cite{ChenCVPR17ordinal} add the $M\!-\!1$ binary class predictions in $\{0,1\}$ to produce an ordinal class in $\{0,M\!-\!1\}$.

In our application, with four ordinal classes, we must train and aggregate three binary classifiers: $\phi_{>0}$: 0 vs. $(1,2,3)$, $\phi_{>1}$: $(0,1)$ vs. $(2,3)$, $\phi_{>2}$: $(0,1,2)$ vs. $3$, where $\phi_{>m}$ is a model to decide whether the target class is greater than value $m$. This approach is appealing because MIL is already well-defined for binary classifications and $\phi_{>0}$ and $\phi_{>1}$ are already relevant tasks introduced in Sec.~\ref{sec:mil}. Thus, it only remains to repeat the MIL method for $\phi_{>2}$.  In the MIL setting, however, the output of each binary classifier results from the frame with the maximum score. Since each classifier in the ensemble is applied to the data independently, the frame with the maximum score may be different for each classifier. We therefore, must first aggregate frame-level binary outputs to produce frame-level ordinal scores and then take the frame with the maximum ordinal score as the representative for the video.

Let $p^i_{>0} = \phi_{>0}(f_i)$, $p^i_{>1} = \phi_{>1}(f_i)$, and $p^i_{>2} = \phi_{>2}(f_i)$ be the ranked binary probability scores in $[0,1]$ for each frame and let $\hat{y}_{>0}^i$, $\hat{y}_{>1}^i$, and $\hat{y}_{>2}^i$ be binary class predictions $\{0,1\}$ for each frame.  We experiment with three ways to aggregate these probabilities to estimate discrete ordinal classes, $\hat{z}^i \in \{0,1,2,3\}$, for each frame:

\begin{enumerate}
\item (Convert) As in \cite{Frank01ordinal}, \underline{convert} the 3 binary probabilities into ordinal class probabilities, $p^i_0 = 1 - p^i_{>0}$, $p^i_1= p^i_{>0} - p^i_{>1}$, $p^i_2 = p^i_{>1} - p^i_{>2}$, $p^i_3 = p^i_{>2}$, and take the max multi-class probability as the ordinal class per frame, $\hat{z}^i = \max_{0\leq j \leq 3} p^i_j$.
\item (Threshold) As in \cite{ChenCVPR17ordinal}, first \underline{threshold} each binary probability score $p^i_{>0}, p^i_{>1}, p^i_{>2}$ with three independent thresholds to produce binary classes predictions $\hat{y}_{>0}^i, \hat{y}_{>1}^i, \hat{y}_{>2}^i$ per frame. Then sum these class predictions to estimate the ordinal class per frame $\hat{z}^i = \hat{y}_{>0}^i + \hat{y}_{>1}^i + \hat{y}_2^i$. 
\item (Sum) First \underline{sum} the three binary probabilities together to produce a continuous severity score $q^i = p^i_{>0} + p^i_{>1} + p^i_{>2} \in [0,3]$. Then apply three ordinal thresholds $0\!<\!t_0\!<\!t_1\!<\!t_2\!<\!3$ to bin the continuous scores into ordinal classes $\hat{z}^i$ per frame.
\end{enumerate} 
Then, to mimic the clinical scoring system, we take the max ordinal class over all frames as the video-level MES severity class estimation, $\hat{z} = \max_{1\leq i\leq N} \hat{z}^i$.  While the first two methods (Convert and Threshold) produce discrete MES severity classes per frame, the third method (Sum) also produces a continuous severity score in [0,3] per frame which may be useful for generating a more precise severity scoring system. Fig.~\ref{fig:ordinalmil} illustrates the proposed ordinal MIL using an ensemble of ranked binary classifiers and the three different aggregation methods.

\subsubsection{Regression with Discrete Labels}
\label{sec:regression}
Another common technique for ordinal classification is to apply regression while maintaining discrete class labels. Here the only difference between classification and regression using CNNs is the final output layer. Instead of a sigmoid activation for binary classification, the final layer is linear with one neuron. Let $\phi_r$ be this regression CNN. Therefore, instead of a binary probability score $p_i\in [0,1]$ as above, this network produces a continuous severity score $s_i \in \mathbb{R}$ per frame such that $s_i\!=\!\phi_r (f_i)$. A clipping function is used to restrict the range of values to [0,3].

\begin{figure}[t]
\centering
\includegraphics[trim={0 10 0 10}, clip, width=\textwidth]{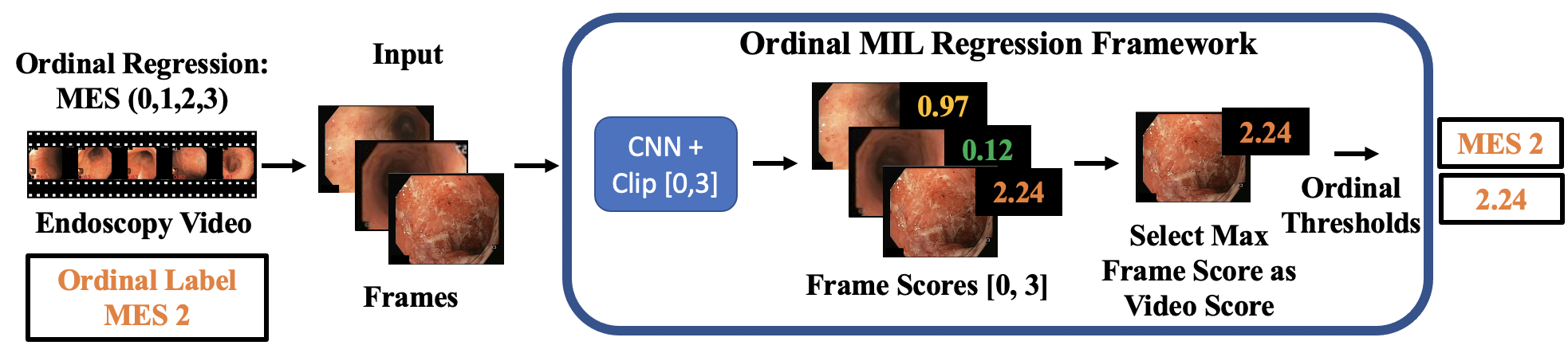}
    \caption{Ordinal MIL Regression framework overview. Endoscopy videos are split into frames and fed into a CNN-based regression model, which generates continuous frame level severity scores in [0,3]. The frame with the maximum score is selected as the representative score for the video. A set of three ordinal thresholds are chosen to bin the continuous scores into discrete MES $\{0,1,2,3\}$. }\label{fig:ordinalmilregression}
\end{figure}

To combine ordinal regression with MIL, we simply take the frame with the maximum severity score, $f_v$, as the severity score as the representative for the video, i.e. $s_v\!=\!\phi_r(f_v)$ with $v\!=\!\argmax_i s_i$. We then apply three ordinal thresholds $0\!<\!t_0\!<\!t_1\!<\!t_2\!<\!3$ to bin the continuous scores into ordinal classes $\hat{z}^i \in \{0,1,2,3\}$ per frame. Fig.~\ref{fig:ordinalmilregression} gives an overview of the ordinal MIL regression method. 
In the next section we compare each of the proposed ordinal MIL methods on clinical trial endoscopy data and validate them on a sample of labeled video frames.

\section{Experiments}
\label{sec:experiments}
\subsection{Data and Training}
\label{sec:data}
%


The data we use in this work originates from the UNIFI clinical trial of Ustekinumab (STELARA[R]) in UC \citep{sands2019ustekinumab}.
Our dataset consists of 1881 endoscopic videos from 726 subjects with 1 to 4 videos per subject lasting anywhere from 35 seconds to 47 minutes and a mean duration of 10 minutes. We extracted frame images from our video files at a frame rate of 1 frame per second (as in \cite{YaoGE20endo}), resulting in a total of 1,129,188 frames with an average of 600 frames per video. 

An inclusion criterion for trial enrollment is an initial screening MES of 2 or 3. During follow-up visits, many subjects had severity reduction to MES 0 or 1. But at the same time, a subset of subjects did not complete the trial. Therefore, the distribution of classes is imbalanced towards the severe classes. Our video class distribution is 167 (MES 0), 220 (MES 1), 492 (MES 2), and 1,002 (MES 3). This imbalance is more pronounced when data samples are relabeled for the set of ranked binary classifiers, i.e., for $\phi_{>0}$: 167 (0), 1,714 (1), $\phi_{>1}$: 387 (0), 1,494 (1), $\phi_{>2}$: 879 (0), 1,002 (1).

To address the class imbalance that skews towards the positive class, we propose a method that exploits the MIL assumption that all frames in a class 0 video must be class 0. Then, instead of selecting the maximum frame that represents a class 0 video, we can select the top $K$ frames \citep{CampanellaNature19path}. This is akin to increasing the sample size $K$ times for the 0 class. While it may be tempting to choose a very large $K$, or all frames from each class 0 video, this may lead to a reverse class imbalance.  
We experimented with $K\!=\!1, 5, 10, 20, 40, 100,$ and $200$ for class 0. For class 1 videos, we leave $K\!=\!1$ since, by construction of MIL, there is no guarantee that more than one frame will exhibit the positive class. 
(If a class 0 video has less than $K$ frames, we select all the frames in that video.) Only the maximum frame is selected at test time. 



After experimentation with various CNN architectures, ResNet34 was chosen as the best performing. The frame image dimensions were resized from $640\!\times\!510\!\times\!3$ to $224\!\times\!224\!\times\!3$ to fit the input to the CNN. After exhaustive parameter testing, we chose an Adam optimizer with learning rate=1e-5, weight decay=0.01, and binary cross entropy loss for 100 epochs for the binary classification tasks. For the ordinal MIL regression we experimented with multiple loss functions including Mean Square Error, Mean Absolute Error, Smooth L1, and Log Cosh and found that Mean Absolute Error provided the best results. We employed online image augmentation of horizontal and vertical flips, rotations, and hue changes.  We ran 5-fold cross validation with splits grouped at the subject-level with proportional label distributions. 

\begin{figure}[t]
    \centering
       \subfloat[Red wall]{\includegraphics[trim={0 0 0 0}, clip, width=0.2375\textwidth]{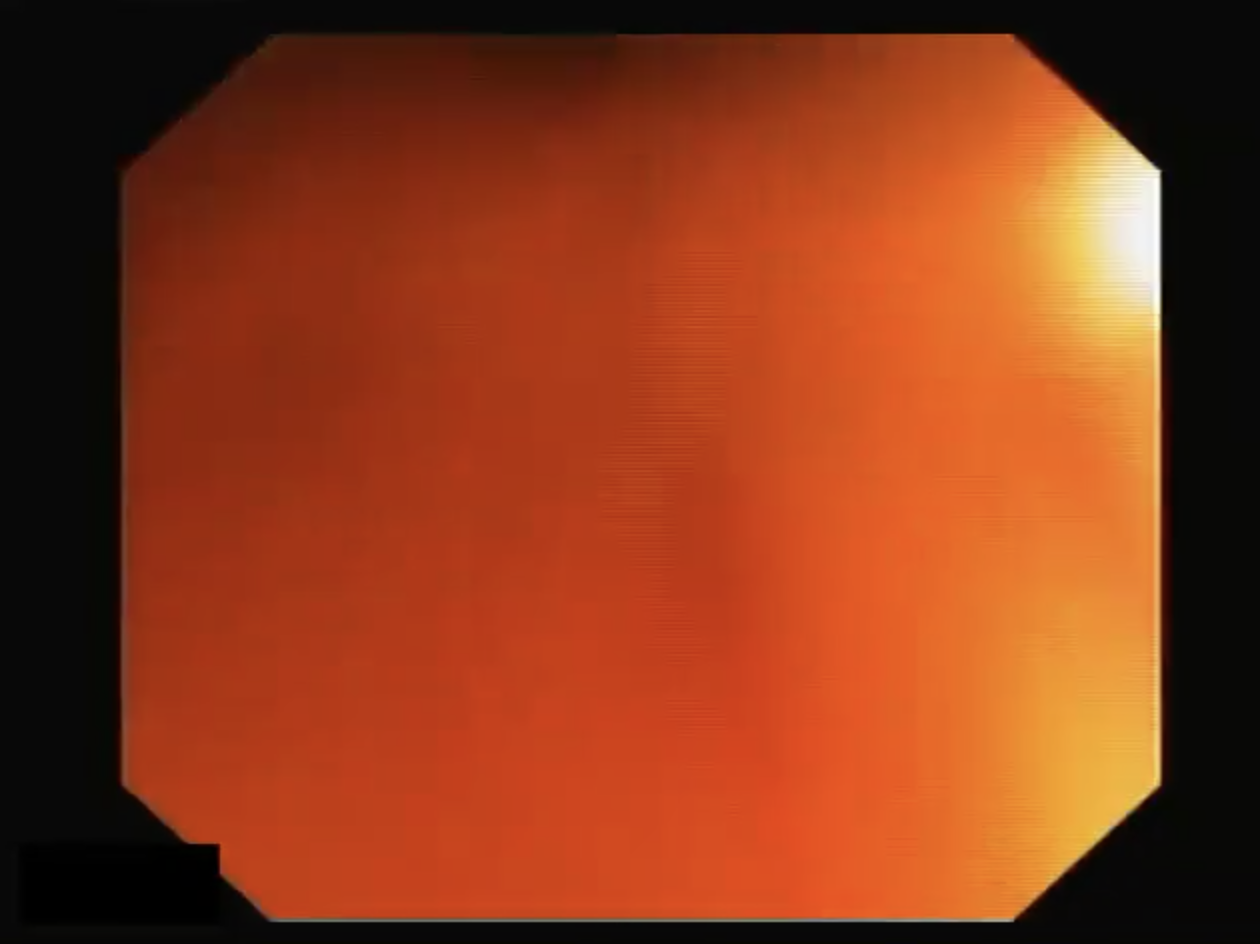}\label{fig:redwall}}
   \subfloat[Bleeding]{\includegraphics[trim={0 0 0 0}, clip, width=0.2395\textwidth]{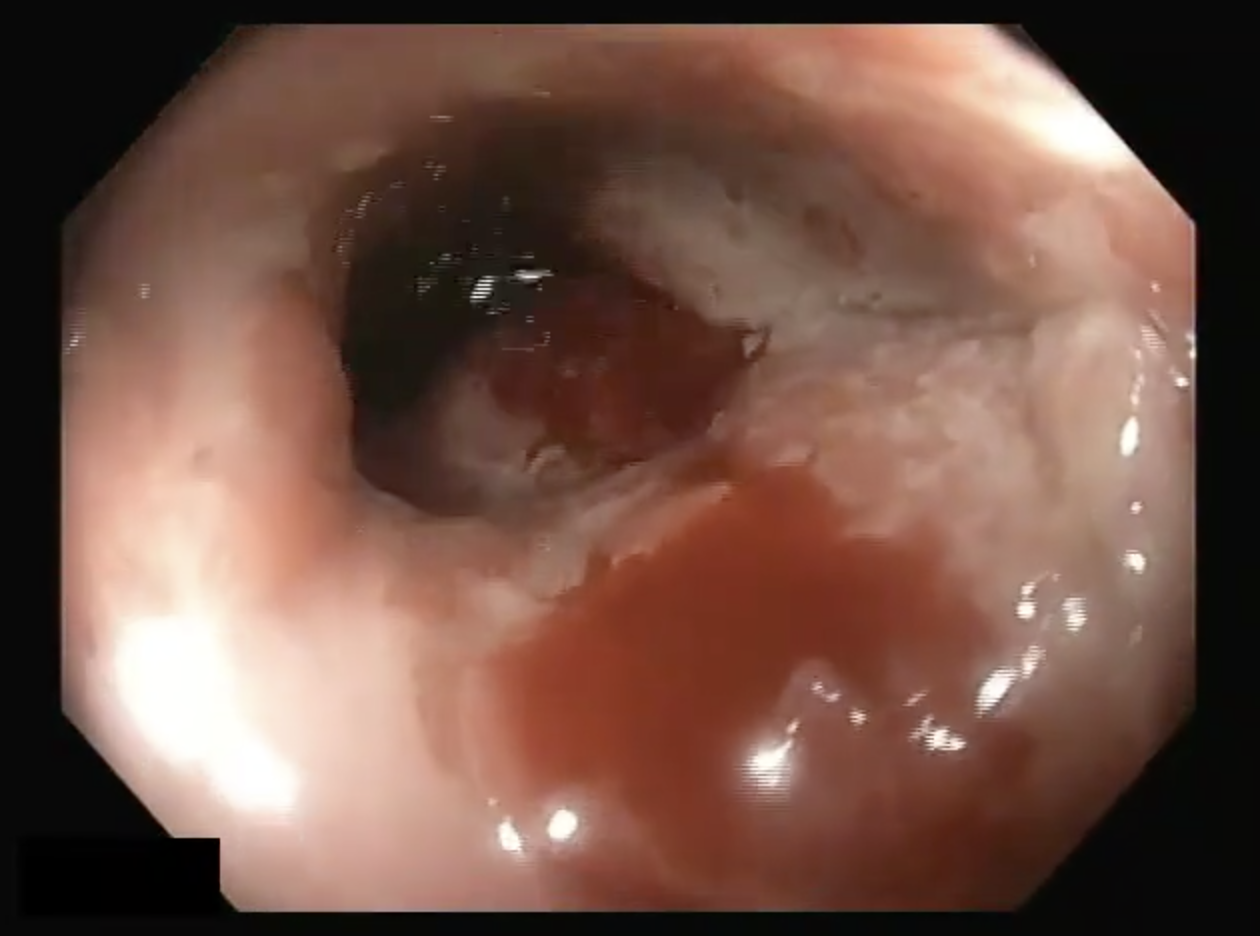}\label{fig:bleeding}}
    \subfloat[Forceps]{\includegraphics[trim={0 0 0 3.5}, clip, width=0.237\textwidth]{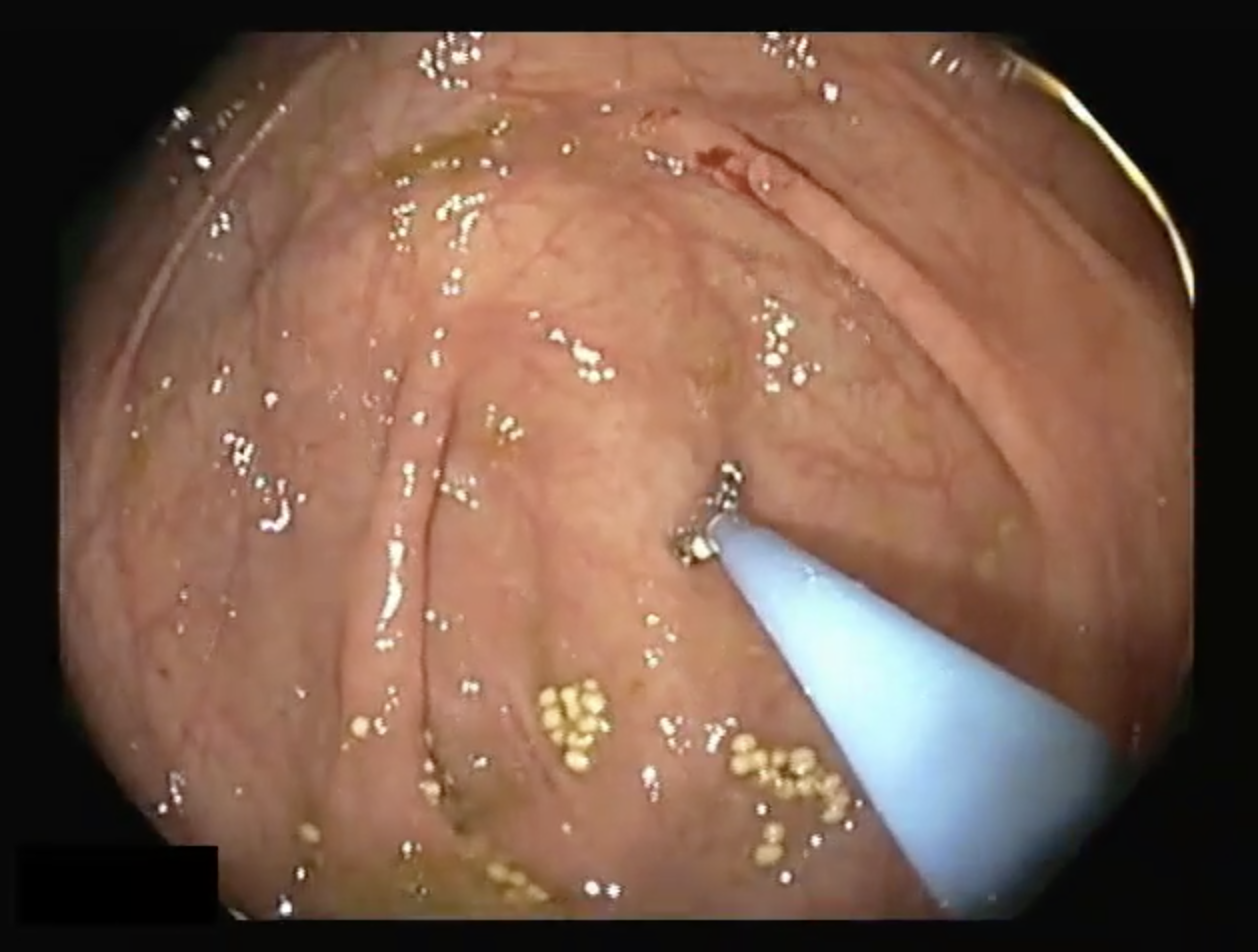}\label{fig:forceps}}
         \subfloat[Rinse Water]{\includegraphics[trim={0 0 0 0}, clip, width=0.239\textwidth]{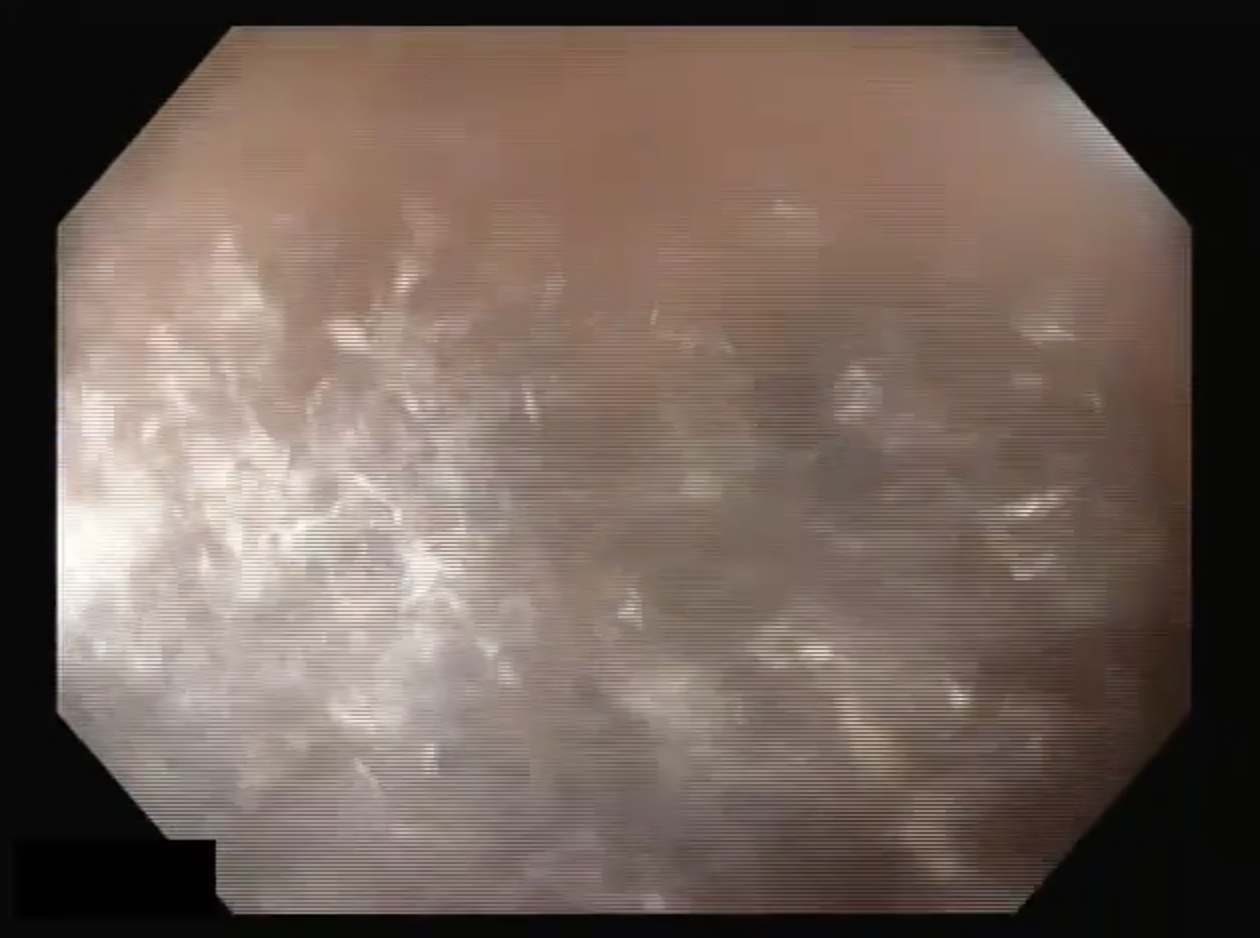}\label{fig:water}}
 \caption{Example endoscopic imaging artifacts. Frames with (a) red wall and (b) bleeding were removed from the dataset due to the similarity with severe inflammation that led to false positive predictions. Frames with (c) forceps and (d) rinse water were not removed since they did not contribute to severe disease  scores.}
 \label{fig:artefacts}
\end{figure}

\subsection{Quality Control}
\label{sec:preprocessing}

Due to the transparent nature of MIL, we ensured a level of quality control by visually inspecting the top scoring frames of true and false positive predictions for consistent image features that may have misled the algorithm. We found that frames that were entirely red (``red wall") (Fig.~\ref{fig:redwall}), where the camera is pressed against the colon, were consistently the highest scoring instead of severe UC. We thus sought to remove these misclassified frames from our dataset to improve performance. 

Taking a supervised approach, we manually labeled a set of 4,200 frames from several videos as being a red wall or not. We used a ResNet pre-trained on ImageNet to extract a feature vector of length 512 from each frame and trained a support vector machine (SVM), achieving $90\%$ accuracy on a validation set of frames from videos without overlapping of subjects. We applied the classifier to the entire dataset and removed 273,262 red wall frames. After successfully removing these red wall frames, we then discovered that images with bleeding (Fig.~\ref{fig:bleeding}) became the highest scoring frames for both true and false positives. We repeated this approach to remove images of bleeding by manually labeling 4,604 images and training a new SVM. With $92\%$ accuracy on a validation set, we removed 222,906 more frames. 

Following the removal of these two types of image artifacts, performance improved allowing severe images to score highly and drive the optimization. We found it was not necessary to remove frames with objects such as forceps\footnote{As per clinical trial protocol, biopsies were taken with forceps at a predefined location in the left colon for all patients and therefore did not correlate to frames with severe disease.} (Fig.~\ref{fig:forceps}) or artifacts like rinse water (Fig.~\ref{fig:water}) because these artifacts did not contribute to severe disease scores. The two rounds of quality control led to a total removal of 496,168 frames, a $32.6\%$ average decrease in the number of frames per video.

\begin{figure}
    \centering
        \includegraphics[trim={10 30 15 0}, clip, width=\textwidth]{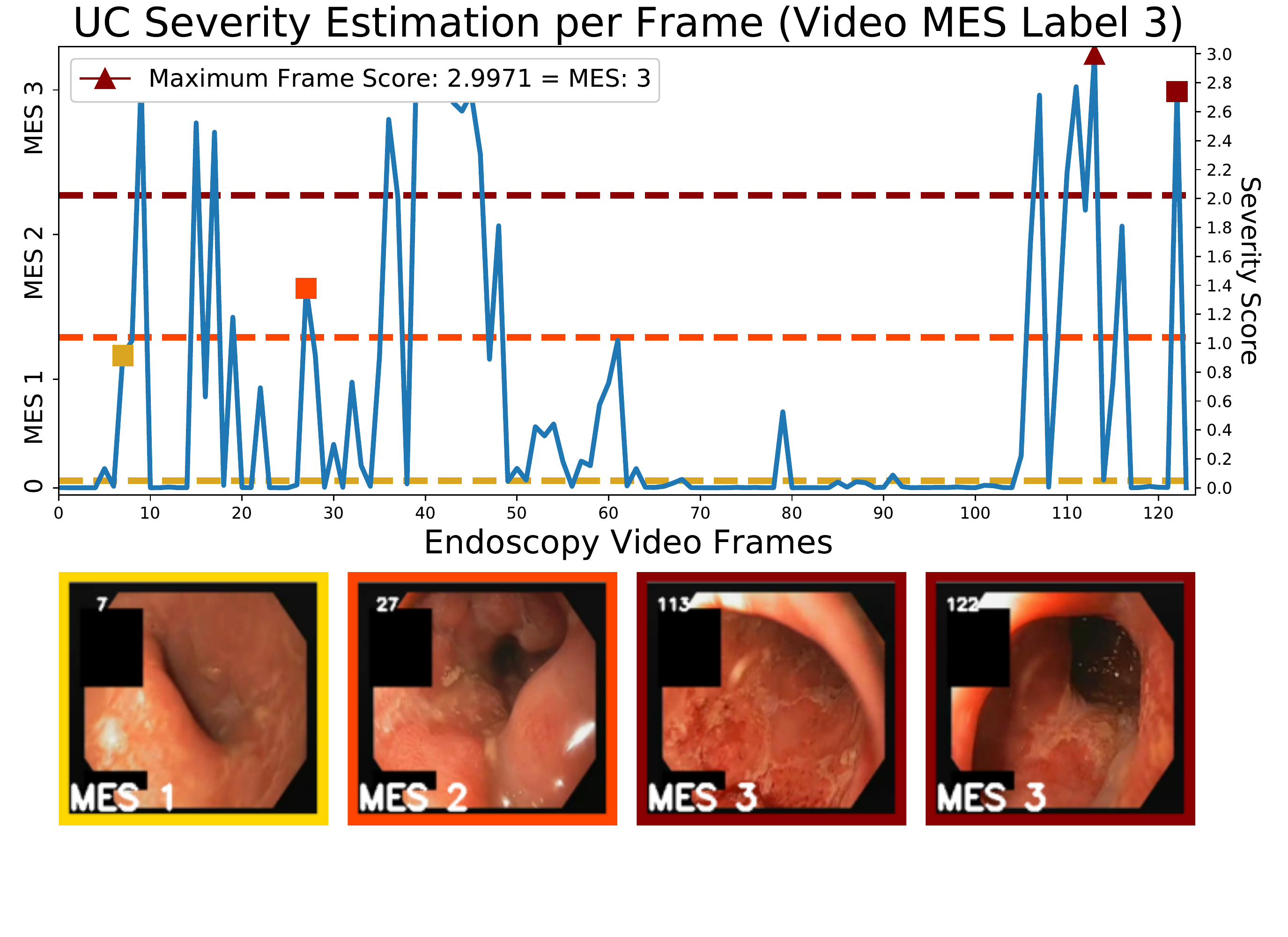}
        \includegraphics[trim={10 40 15 0}, clip, width=\textwidth]{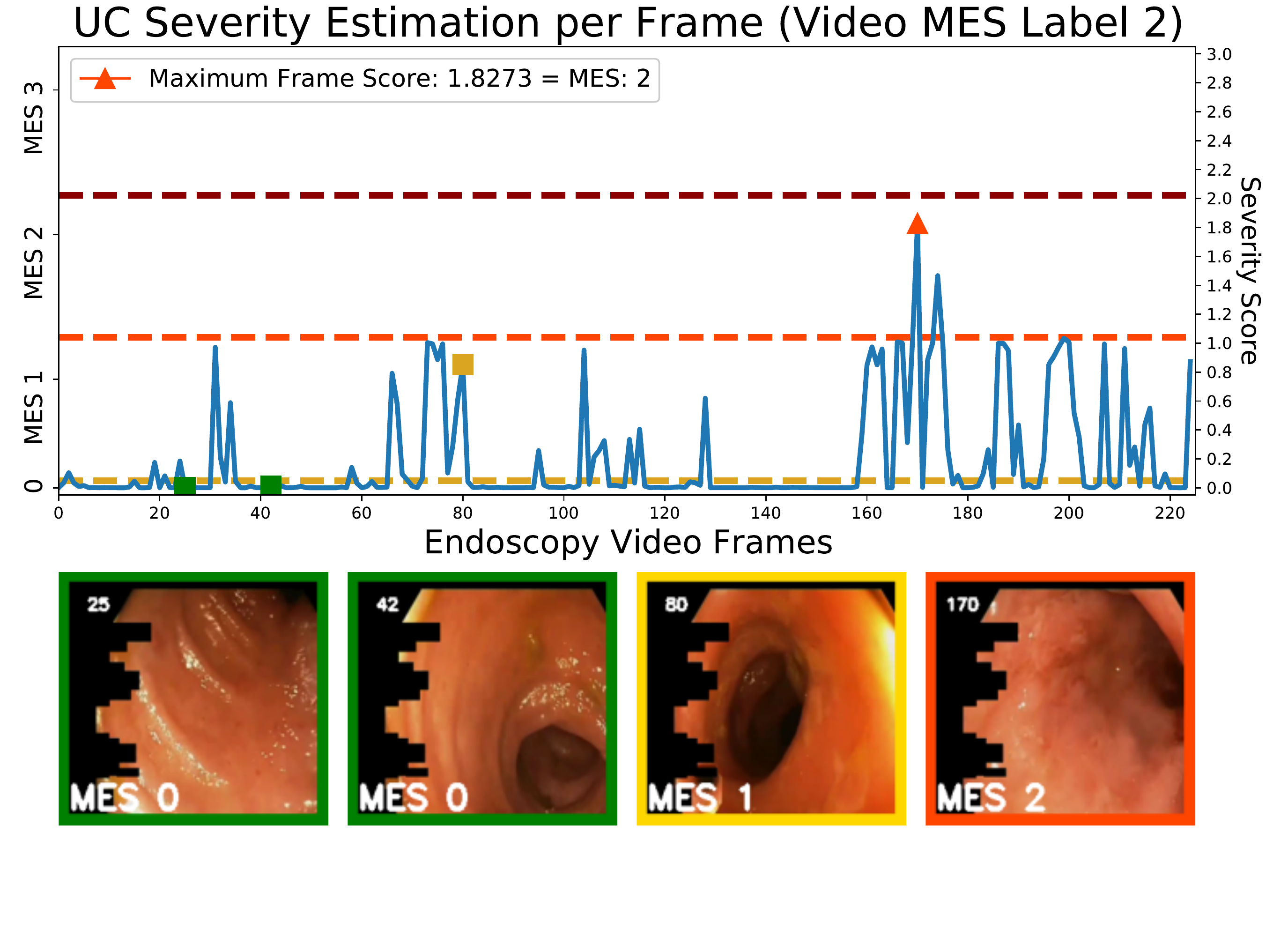}
        \caption{Continuous frame-level MES severity estimations for two endoscopy videos using the ordinal MIL ensemble method 3 (Sum).  Each sample frame corresponds to the square colored markers on the plot, and the maximum frame, indicated by a triangle marker, gives the score for the video. Each dotted line is an optimal MES threshold found via grid search over 5-fold cross validation.}
        \label{fig:severityplots}
\end{figure}

\begin{figure}[h]
\centering
\includegraphics[trim={0 0 0 0}, clip, height=.4\textwidth, width=.7\textwidth]{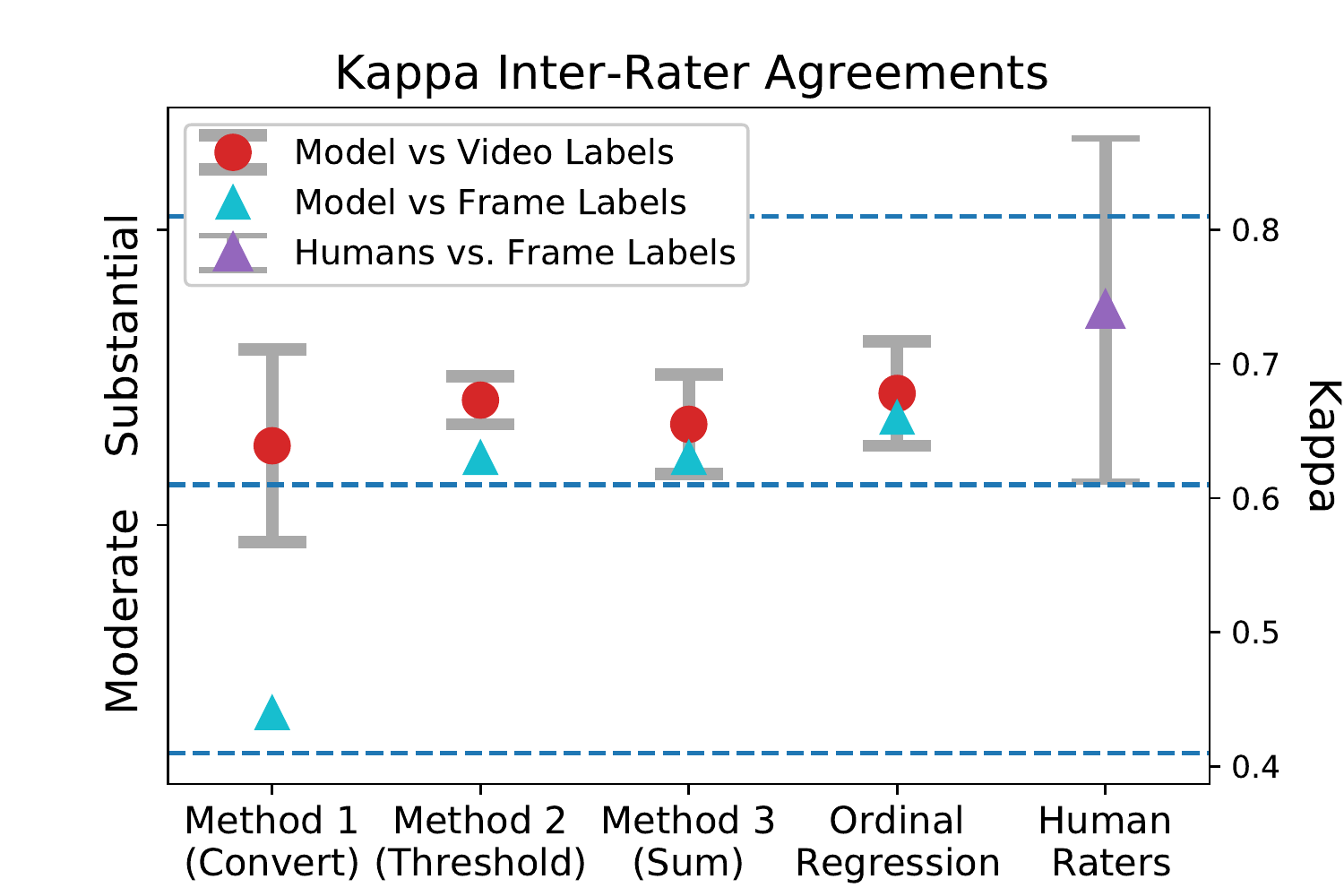}
\includegraphics[trim={0 0 0 0}, clip, width=.45\textwidth]{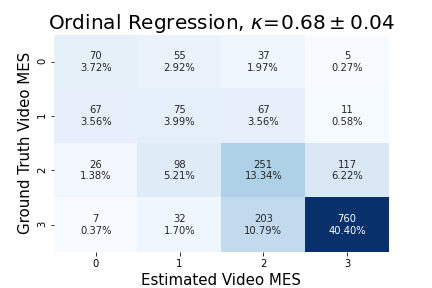}
\includegraphics[trim={0 0 0 0}, clip, width=.45\textwidth]{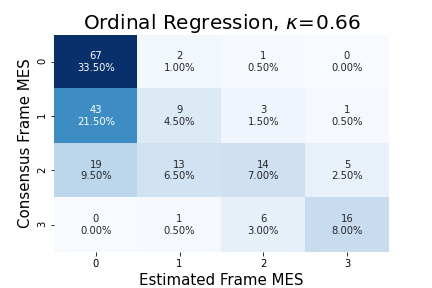}
\caption{Comparison of inter-rater agreement measured by quadratic weighted Cohen's $\kappa$ between the MES estimations of each of the proposed methods and MES labels.  Red circles (with $95\%$ confidence intervals (CI) over 5-fold cross validation) indicate the agreement between model estimations and video-level labels. Each method achieves substantial agreement with ground truth labels. Blue triangles indicate the agreement between model estimations and frame-level consensus labels for 200 frames within a validation split. Purple triangle is the average inter-rater agreement between each human rater and the frame-level consensus labels with $95\%$ CI. 
The confusion matrices show MES estimations vs. ground truth for Ordinal MIL Regression for video and frame-level labels. The video-level confusion matrix includes estimations over all 5-folds.}
\label{fig:kappaconmats}
\end{figure}


\section{Results}
\label{sec:results}
\subsection{Video-Level Analysis}
\label{sec:videolevelanalysis}
Our first results are for the three ranked binary classification tasks. 
With $K=\{40,200,100\}$ we achieved AUCs of $0.90\pm 0.03$, $0.92\pm 0.02$, and $0.88\pm 0.02$ for $\phi_{>0}$, $\phi_{>1}$, and $\phi_{>2}$, respectively, across 5-fold cross validation. While $K$ parameters were tuned to achieve these AUCs for each classifier independently, we found that parameters may also be tuned with respect to their combination for ordinal classification. The combination of $K\!=\!\{40,100,40\}$ gave us the best average ordinal classification performance over the 5-fold cross validation for each aggregation method in Sec.~\ref{sec:ensemble}.

For the ordinal ensemble model 2 (Threshold), binary thresholds are needed for each of the three binary classifiers. Using a grid search over 5-fold cross validation, we selected the optimal combination of binary thresholds $\{t_1,t_2,t_3\}\!=\!\{0.20,0.19,0.22\}$. For the ordinal ensemble model 3 (Sum) which produces continuous severity scores ranging from [0,3], ordinal thresholds are needed to bin these scores into discrete MES.  Using a grid search over 5-fold cross validation, we selected ordinal thresholds $\{t_1,t_2,t_3\}\!=\!\{0.1,1.04,2.02\}$ to maximize performance. Fig.~\ref{fig:severityplots} illustrates our frame-level severity estimation for two example videos from a validation set using the ensemble summation method 3 to produce continuous severity scores.  The frames with the maximum score correctly estimate the ground truth MES. For the ordinal MIL regression method, we found $K\!=\!10$ produced the best performance. Over 5-fold cross validation we found optimal ordinal thresholds $\{t_1,t_2,t_3\}\!=\!\{1.03,1.97,2.79\}$.


In evaluating the performances of our methods, we calculated a quadratic weighted Cohen's $\kappa$ value between the output of our models and the ground truth labels.  Because the ground truth have been shown to be subjective and variable as indicated by a moderate Fleiss\footnote{The (unweighted) Fleiss $\kappa$ reported by studies including our own data, are not necessarily comparable to a weighted Cohen's $\kappa$, so we do not make claims of comparison to human raters at the video-level.} $\kappa$, measuring accuracy may be a misleading quantity.  Instead, we aim to demonstrate that we achieve an acceptable level of agreement with the ground truth. Our three ordinal MIL ensemble methods achieved substantial agreement with $\kappa\!=\!0.64$ ($95\%$ CI 0.56-0.71), $\kappa\!=\!0.67$ ($95\%$ CI 0.66-0.69), and $\kappa\!=\!0.65$ ($95\%$ CI 0.62-0.69), respectively. The ordinal MIL regression model also achieved a substantial agreement with $\kappa\!=\!0.68$ ($95\%$ CI  0.64-0.72). Fig.~\ref{fig:kappaconmats} shows a plot of the $\kappa$ comparisons and confusion matrices for the ordinal MIL regression model. These results show a very similar performance between all methods at the video-level. In the next section, we evaluate performance at the frame-level.

\subsection{Frame-Level Analysis}
\label{sec:framelevelanalysis}

While the above analyses show substantial $\kappa$ values at the video level, it is also important that we also validate our methods at the frame-level. To this end, we collected frame-level MES labels from 4 experienced, non-central reader GIs to compute inter-rater agreement at the frame-level. We randomly selected 50 frames from videos of each of the 4 MES classes within a single validation split of our dataset, resulting in 200 total frames. This task was time consuming, requiring 3-4 hours per physician to label all 200 frames. To compare the inter-rater agreement of the 4 raters, we calculated a Fleiss $\kappa\!=\!0.42$, which is on the low end of moderate and significantly below the reported video-level Fleiss $\kappa\!=\!0.60$ ($95\%$ CI 0.51-0.69) \citep{sands2019ustekinumab}.

In addition, because we know the ground truth video labels from which each of the 200 frames were extracted, we analyzed how well the frame-level labels coincided with the video-level labels using the fact that the frame MES label should not exceed the video MES label. In fact, $27.5\%$ of the annotated frame labels were greater than the video scores. We adjusted the frame labels to be consistent with the video labels by taking the minimum of the two and then took a majority vote consensus to produce our frame labels. In addition to a Fleiss $\kappa$, we calculated how well each rater agreed with the adjusted consensus frame labels by calculating the quadratic weighted Cohen's $\kappa\!=\!0.80, 0.78, 0.77, 0.62$ with an average $\kappa\!=\!0.74$ (CI 0.61-0.87).

In comparison, for each of our models we achieved quadratic weighted Cohen's $\kappa\!=\!0.44$ (Convert) $\kappa\!=\!0.63$ (Threshold), $\kappa\!=\!0.63$ (Sum), and $\kappa\!=\!0.66$ (Ordinal MIL Regression), respectively.  We found that the Threshold, Sum and the ordinal MIL regression models achieve a substantial $\kappa$ value that falls within the $95\%$ CI of the human raters at the frame-level. See Fig.~\ref{fig:kappaconmats} for this $\kappa$ comparison and accompanying confusion matrices for the ordinal MIL regression. These results indicate that the proposed ordinal MIL methods, can not only automate the established MES scoring system at the video level with substantial agreement to the ground truth, but can also provide frame-level MES estimations comparable to human raters.

\section{Conclusion}
\label{sec:conclusion}

In this work, we proposed a set of novel weakly supervised ordinal classification methods to automatically estimate the severity of UC from endoscopy videos.  Our methods have the ability to estimate frame-level severity on a continuous scale by training only on discrete video-level severity labels.  This is beneficial for expanding models quickly to other disease types and clinical indications without expensive, time-consuming and subjective labeling for each new target of interest. Our proposed ordinal MIL framework has the potential to effectively automate the established video-level MES scoring system with a substantial level of score agreement. Furthermore, we validate that the resulting frame-level severity estimations are comparable to that of human raters. With continuous severity scores at the frame level, our method will foster the development of more localized and granular endoscopic endpoints that may be more sensitive to meaningful therapeutic effects and better predict patient outcomes in IBD.

\section*{Acknowledgement}
We give special thanks to Dr. Louis Ghanem, Dr. Kathleen Lomax, Dr. Laurie Conklin, and Dr. Susana Gonzalez for providing their GI expertise in scoring a set of endoscopy video frames for use in our method validation and analysis.

%
%


%
%
%
%
\end{document}